\documentclass[prd,twocolumn,showpacs,floatfix]{revtex4}
\usepackage{amsmath,amssymb,graphicx}

\begin{document}

\title{Stability of the Einstein static universe in
modified Gauss-Bonnet gravity}

\author{Christian G. B\"ohmer}
\email{c.boehmer@ucl.ac.uk}
\affiliation{Department of Mathematics and Institute of Origins,
University College London, Gower Street, London, WC1E 6BT, UK}

\author{Francisco S. N. Lobo}
\email{flobo@cosmo.fis.fc.ul.pt}
\affiliation{Centro de Astronomia e Astrof\'{\i}sica
da Universidade de Lisboa, Campo Grande,
Ed. C8 1749-016 Lisboa, Portugal}

\date{\today}

\begin{abstract}
We analyze the stability of the Einstein static universe by
considering linear homogeneous perturbations in the context of $f({\cal G})$
modified Gauss-Bonnet theories of gravity. By considering a
generic form of $f({\cal G})$, the stability region of the Einstein
static universe is parameterized by the linear equation of state
parameter $w=p/\rho$ and the second derivative $f''({\cal G})$ of the
Gauss-Bonnet term.
\end{abstract}

\pacs{04.50.+h, 04.20.Jb, 04.25.Nx}
\maketitle

\section{Introduction}

Cosmology is said to be thriving in a golden age, where a central
theme is the perplexing fact that the Universe is undergoing an
accelerating expansion~\cite{expansion}. The latter, one of the
most important and challenging current problems in cosmology,
represents a new imbalance in the governing gravitational
equations. Historically, physics has addressed such imbalances by
either identifying sources that were previously unaccounted for,
or by altering the governing equations. The cause of this
acceleration still remains an open and tantalizing question.
Although the introduction of a cosmological constant into the
field equations seems to be the simplest theoretical approach to
generate a phase of accelerated expansion, several alternative
candidates have been proposed in the literature, ranging from
dynamical dark energy models to modified theories of gravity.
Amongst the latter, models generalizing the Einstein-Hilbert
action have been proposed.

The Einstein field equations of General Relativity were first
derived from an action principle by Hilbert, by adopting a linear
functional of the scalar curvature, $R$, in the gravitational
Lagrangian density. However, there are no \textit{a priori} reasons
to restrict the gravitational Lagrangian to this form, and indeed
several generalizations of the Einstein-Hilbert Lagrangian have
been proposed, including ``quadratic Lagrangians'', involving
second order curvature invariants such as $R^{2}$, $R_{\mu
\nu}R^{\mu \nu }$, $R_{\alpha \beta \mu \nu }R^{\alpha \beta \mu
\nu}$, $\varepsilon ^{\alpha \beta \mu \nu }R_{\alpha \beta \gamma
\delta }R_{\mu \nu }^{\gamma \delta }$, $C_{\alpha \beta \mu
\nu}C^{{\alpha \beta \mu \nu }}$, etc~\cite{early} (see Ref.~%
\cite{Lobo:2008sg} for a recent review). The physical motivations
for these modifications of gravity were related to the possibility
of a more realistic representation of the gravitational fields
near curvature singularities and to create some first order
approximation for the quantum theory of gravitational fields.

In considering alternative higher-order gravity theories, one is
liable to be motivated in pursuing models consistent and inspired
by several candidates of a fundamental theory of quantum gravity.
In this context, it may be possible that unusual gravity-matter
couplings predicted by string/M-theory may become important at the
recent low-curvature Universe. For instance, one may couple a
scalar field not only with the curvature scalar, as in
scalar-tensor theories, but also with higher order curvature
invariants. Indeed, motivations from string/M-theory predict that
scalar field couplings with the Gauss-Bonnet invariant ${\cal G}$
are important in the appearance of non-singular early time
cosmologies \cite{Antoniadis:1993jc}. It is also possible to apply
these motivations to the late-time Universe in an effective
Gauss-Bonnet dark energy model \cite{Nojiri:2005vv}.

In the context of $f({\cal G})$ modified theories of gravity, we
explore the stability of the Einstein static universe in this
work. This is motivated by the possibility that the universe might
have started out in an asymptotically Einstein static state, in
the inflationary universe context \cite{Ellis:2002we}. On the
other hand, the Einstein cosmos has always been of great interest
in various gravitational theories. In general relativity for
instance, generalizations with non-constant pressure have been
analyzed in \cite{ESa}. In brane world models, the Einstein static
universe was investigated in \cite{Gergely:2001tn}, while its
generalization within Einstein-Cartan theory can be found in
\cite{Boehmer:2003iv}, and in loop quantum cosmology, we refer the
reader to \cite{Mulryne:2005ef}.

In the context of $f(R)$ modified theories of gravity, the
stability of the Einstein static universe was also analyzed by
considering homogeneous perturbations \cite{Boehmer:2007tr}. By
considering specific forms of $f(R)$, the stability regions of the
solutions were parameterized by a linear equation of state
parameter $w=p/\rho$. Contrary to classical general relativity, it
was found that in $f(R)$ gravity a stable Einstein cosmos with a
positive cosmological constant does indeed exist. Thus, in
principle, modifications in $f(R)$ gravity stabilize solutions
which are unstable in general relativity. Furthermore, in
\cite{Goswami:2008fs} it was found that only one class of $f(R)$
theories admits an Einstein static model, and that this class is
neutrally stable with respect to vector and tensor perturbations
for all equations of state on all scales. These results are
apparently contradictory with those of Ref.~\cite{Boehmer:2007tr}.
However, in a recent work, homogeneous and inhomogeneous scalar
perturbations in the Einstein static solutions were analyzed
\cite{Seahra:2009ft}, consequently reconciling both of the above
works.

This paper is outlined in the following manner: In Section
\ref{sec:II}, we briefly review modified Gauss-Bonnet gravity, for
self-completeness and self-consistency, and present the respective
field equations. In Section \ref{sec:III}, we consider a generic
form of $f({\cal G})$, and analyze the stability of the solutions,
by considering homogeneous perturbations around the Einstein
static universe, and deduce the respective stability regions. The
latter are given in terms of the linear equation of state
parameter $w=p/\rho$ and the unperturbed energy density $\rho_0$.
Finally, in Section \ref{sec:concl} we present our conclusions.
Throughout this work, we consider the following units $c=G=1$.

\section{Modified Gauss-Bonnet gravity and field equations}
\label{sec:II}


An interesting alternative gravitational theory is modified
Gauss-Bonnet gravity, which is given by the following action
\begin{align}
  S=\int d^4x \sqrt{-g} \left[\frac{R}{2\kappa}+f({\cal G})\right]
  + S_M(g_{\mu\nu},\psi)\,,
  \label{modGBaction}
\end{align}
where $\kappa = 8\pi$. $f({\cal G})$ is an arbitrary function of
the Gauss-Bonnet invariant, which is in turn given by
\begin{align}
  {\cal G}\equiv
  R^2-4R_{\mu\nu}R^{\mu\nu}+R_{\mu\nu\alpha\beta}R^{\mu\nu\alpha\beta}\,.
  \label{modGBinv}
\end{align}

$S_M(g_{\mu\nu},\psi)$ is the matter action, defined as $S_M=\int
d^4x\sqrt{-g}\;{\cal L}_m(g_{\mu\nu},\psi)$, where ${\cal L}_m$ is
the matter Lagrangian density, in which matter is minimally
coupled to the metric $g_{\mu\nu}$ and $\psi$ collectively denotes
the matter fields. The matter stress-energy tensor, $T_{\mu
\nu}^{(m)}$, is defined as $T_{\mu \nu}^{(m)}=-\frac{2}{\sqrt{-g}}
  \frac{\delta(\sqrt{-g}\,{\cal L}_m)}{\delta(g^{\mu\nu})}$.
Thus, using the diffeomorphism invariance of
$S_M(g_{\mu\nu},\psi)$ yields the covariant conservation of the
stress-energy tensor, $\nabla^\mu T_{\mu\nu}^{(m)}=0$.

Let us for the moment assume the function $f({\cal G})$ to be
expanded around the origin so that we may write
\begin{align}
  f({\cal G}) = f(0) + f'(0){\cal G} + \frac{f''(0)}{2!}{\cal G}^2
  + O({\cal G}^3)\,.
\end{align}
If we insert this representation back into the Lagrangian, we note
that $f(0)$ acts like an effective cosmological constant. The term
$f'(0)$ would not be present in the field equations because this
term is linear in the Gauss-Bonnet invariant and linear
topological terms do not contribute to the equations of motion.
The first non-trivial term to be expected is $f''(0)$. Moreover,
as we are interested in linear perturbation theory, we expect only
the term $f''(0)$ to characterize deviations from general
relativity. Hence, we anticipate that the general relativistic
limit takes the form $f''(0) \rightarrow 0$.

Modified Gauss-Bonnet gravity has been extensively analyzed in the
literature, and rather than review all of its intricate details
here, we refer the reader to
Refs.~\cite{modGB1,modGB2,Nojiri:2005jg}. Varying the action
Eq.~(\ref{modGBaction}) with respect to $g^{\mu\nu}$, one obtains
the gravitational field equations in the following form
$G^{\mu\nu}\equiv R^{\mu\nu}-\frac{1}{2}g^{\mu\nu}R=\kappa\,
  T^{\mu\nu}_{\rm eff}$,
where the effective stress-energy tensor is defined as
$T^{\mu\nu}_{\rm eff}=T^{\mu\nu}_{(m)}+T^{\mu\nu}_{\rm GB}$. The
effective Gauss-Bonnet curvature stress-energy term is given by
\begin{widetext}

\vspace{-0.5cm}

\begin{align}
  T^{\mu\nu}_{\rm GB}&=\frac{1}{2}g^{\mu\nu}f({\cal G})
  -2f'({\cal G})RR^{\mu\nu}+4f'({\cal G})R^{\mu}{}_{\rho}R^{\nu\rho}
  -2f'({\cal G})R^{\mu\rho\sigma\tau}R^{\nu}{}_{\rho\sigma\tau}
  -4f'({\cal G})R^{\mu\rho\sigma\nu}R_{\rho\sigma}+2[\nabla^\mu
  \nabla^\nu f'({\cal G})] R
  \nonumber \\
  &-2g^{\mu\nu}[\nabla^2 f'({\cal G})] R-4[\nabla_\rho \nabla^\mu
  f'({\cal G})]R^{\nu\rho}-4[\nabla_\rho \nabla^\nu f'({\cal G})]
  R^{\mu\rho}+4[\nabla^2 f'({\cal G})] R^{\mu\nu}
  \nonumber \\
  &+4g^{\mu\nu}[\nabla_\rho \nabla_\sigma f'({\cal G})]R^{\rho\sigma}
  -4[\nabla_\rho \nabla_\sigma f'({\cal G})]R^{\mu\rho\nu\sigma}\,,
  \label{modGBeffSET}
\end{align}
\end{widetext}
where the definition $f'({\cal G}) \equiv df({\cal G})/d{\cal G}$
is considered for notational simplicity.

\section{The Einstein static Universe in $f({\cal G})$
gravity and perturbations} \label{sec:III}

\subsection{Metric and field equations}

Consider the metric given by
\begin{align}
  ds^2=-dt^2+a^2(t)\left[\frac{dr^2}{1-r^2}+r^2\,(d\theta^2
   +\sin^2{\theta} \, d\phi ^2)\right] \,.
  \label{Einst:metric2}
\end{align}
The Gauss-Bonnet invariant in provided by
\begin{equation}
  {\cal G}= 24 \frac{1+a'^2}{a^2} \frac{a''}{a}\,,
  \label{modGBinvESU}
\end{equation}
where the prime, in this context, denotes differentiation with
respect to cosmological time $t$.

For the Einstein static universe, $a=a_0={\rm const}$, the Ricci
scalar becomes $R=6/a_0^2$, and the Gauss-Bonnet invariant is
trivially given by ${\cal G}(a_0) = 0$. The field equations in
this case take the following form
\begin{align}
  \frac{3}{a_0^2} = \kappa \left(\rho_0 - \frac{f(0)}{2}\right)\,,
   \qquad
  -\frac{1}{a_0^2} = \kappa \left(p_0 + \frac{f(0)}{2}\right)\,,
  \label{back:p0}
\end{align}

where $\rho_0$ and $p_0$ are the unperturbed energy density and
isotropic pressure, respectively. It should be noted that the
modified Gauss-Bonnet term $f(0)$ acts like an effective
cosmological constant to the background field equations, as noted
above.

\subsection{Linear homogeneous perturbations}

In what follows, we analyze the stability against linear
homogeneous perturbations around the Einstein static universe
given in Eqs.~(\ref{back:p0}). Thus, we introduce perturbations in
the energy density and the metric scale factor which only depend
on time
\begin{align}
  \rho(t) = \rho_0\left(1+\delta\rho(t)\right),
  \qquad
  a(t) = a_0\left(1+\delta a(t)\right).
  \label{def:perturbations}
\end{align}
Subsequently, we consider a linear equation of state,
$p(t)=w\rho(t)$, linearize the perturbed field equations and
analyze the dynamics of the solutions.

Despite the Gauss-Bonnet invariant being trivial at the background
level, the perturbations have an effect. From
Eq.~(\ref{modGBinvESU}) it follows that
\begin{align}
  \delta {\cal G} = 24 \frac{k}{a_0^2}\frac{\delta a''}{a_0}\,,
\end{align}
in linear order. Therefore, using $f({\cal G} + \delta {\cal G}) =
f({\cal G}) + f'({\cal G})\delta {\cal G}$ plus higher order
terms, we find for linear perturbations
\begin{align}
  f({\cal G}) = f(0) + 24 f'(0) \frac{k}{a_0^2}\frac{\delta a''}{a_0}\,,
\end{align}
where we assumed that $f$ is an analytic function.

We now insert the perturbations~(\ref{def:perturbations}) into the
full field equations and linearize. This provides the following
two equations
\begin{align}
  \frac{6}{a_0^2} \delta a + \kappa \rho_o \delta\rho = 0\\
  \frac{2}{a_0^2} \delta a - 2\delta a'' - \kappa \rho_0 w
  \delta\rho - 96 \kappa \frac{f''(0)}{a_0^4} \delta a^{(4)} = 0
\end{align}
It should be noted that these perturbation equations do not
contain contributions of the form $f'(0)$, as expected. The first
equation relates the perturbations in the scale factor to the
density perturbations. Next, we can eliminate $\delta\rho$ from
the second equation using the first and we arrive at the following
fourth order perturbation equations for the perturbed scale factor
\begin{align}
  \frac{2}{a_0^2}(1+3w)\delta a - 2 \delta a'' -
  96 \kappa \frac{f''(0)}{a_0^4} \delta a^{(4)} = 0\,.
\end{align}
By virtue of the background equations we have
\begin{align}
  \frac{2}{a_0^2} = \kappa (1+w) \rho_0\,,
\end{align}
and therefore we finally arrive at the following perturbation
equation
\begin{align}
  24 \kappa^3 &\rho_0^2 (1+w)^2 f''(0) \delta a^{(4)}(t)
  \nonumber \\&+
  2 \ddot{\delta a}(t) - \kappa \rho_0 (1+w)(1+3w) \delta a(t) = 0\,.
  \label{perteq}
\end{align}

\subsection{Stability regions}

Using the standard ansatz $\delta a(t) = C \exp(\omega t)$, where
$C$ and $\omega$ are constants, we find that Eq.~(\ref{perteq})
provides as solutions the following four frequencies of the small
perturbations
\begin{align}
  \omega_{\pm}^2 = \frac{-1 \pm \sqrt{1 + 24 \kappa^4 \rho_0^3
  (1+w)^3 (1+3w) f''(0)}}{24 \kappa^3 \rho_0^2 (1+w)^2 f''(0)}\,.
\end{align}

As expected, in the limit $f''(0) \rightarrow 0$ the frequencies
$\omega_{+}^2$ lead to the general relativistic result
\begin{align}
  \omega_{\rm GR}^2 = \frac{\kappa\rho_0}{2}(1+w)(1+3w)\,,
\end{align}
while the frequencies $\omega_{-}^2$ become formally infinite in
this limit, which is simply an artifact of the reduction of
fourth-order to second order gravity.

The perturbations $\delta a$ are stable if the frequencies
$\omega$ are purely complex, which implies that the stability is
equivalent to the two conditions $\omega_{\pm}^2 < 0$. In order to
simplify the notation, let us introduce the new parameter $\alpha
= 24 \kappa^4 \rho_0^3 f''(0)$, so that the perturbations are
stable for the following three parameter regions.

First, there is a region analogous to the classical general
relativistic stability region given by
\begin{align}
  -\frac{1}{3} > w > -1\,,
  \qquad -\frac{1}{(1+w)^3(1+3w)} \geq  \alpha > 0\,.
\end{align}

The second region covers the range where the equation of state is
larger due to the presence of the Gauss-Bonnet term, and is
provided by
\begin{align}
  w > -\frac{1}{3}\,,
  \qquad 0 > \alpha \geq -\frac{1}{(1+w)^3(1+3w)}\,.
\end{align}

Finally, there exits a stable region where the equation of state
is less than minus one. We denote this the phantom region, and the
stability regions are given by
\begin{align}
  w < -1 \,,
  \qquad 0 > \alpha \geq -\frac{1}{(1+w)^3(1+3w)}\,.
\end{align}

\begin{figure}[!htb]
\begin{center}
\includegraphics[width=.5\textwidth]{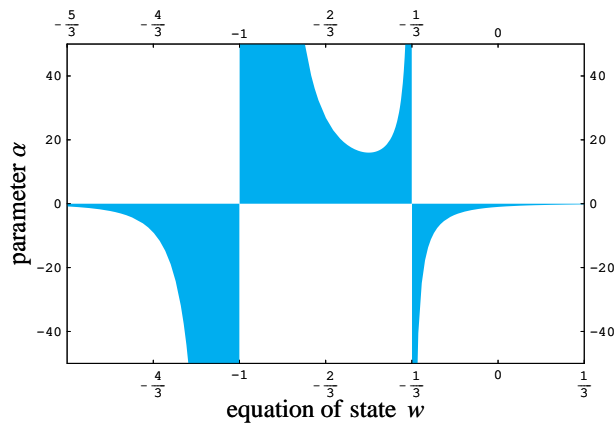}
\end{center}
\caption{Regions of stability in the $(w,\alpha)$ parameter space
for homogeneous perturbations of Einstein static universes. The
left is the phantom region, the middle part is the general
relativistic analogue region and the right part corresponds to a
normal matter region. Note that there exist stable modes for all
$w$, provided $\alpha$ is chosen appropriately.} \label{fig:stab}
\end{figure}

These three regions are depicted in Fig.~\ref{fig:stab} and it is
evident that there exits stable modes for all equation of state
parameters $w$ if $\alpha$ is chosen appropriately. We emphasize
that this includes positive values of the equation of state. It is
interesting to note that the negative values of $\alpha = 24
\kappa^4 \rho_0^3 f''(0)$, imply that $f''(0)<0$.\\

\section{Summary and discussion}
\label{sec:concl}

The Einstein static universe has recently been revived as the
asymptotic origin of an emergent universe, namely, as an
inflationary cosmology without a singularity \cite{Ellis:2002we}.
The role of positive curvature, negligible at late times, is
crucial in the early universe, as it allows these cosmologies to
inflate and later reheat to a hot big-bang epoch. An attractive
feature of these cosmological models is the absence of a
singularity, of an `initial time', of the horizon problem, and the
quantum regime can even be avoided. Furthermore, the Einstein
static universe was found to be neutrally stable against
inhomogeneous linear vector and tensor perturbations, and against
scalar density perturbations provided that the speed of sound
satisfies $c_{\rm s}^2>1/5$ \cite{Barrow:2003ni}. Further issues
related to the stability of the Einstein static universe may be
found in Ref.~\cite{Barrow}.

In this work we have analyzed linear homogeneous perturbations
around the Einstein static universe in the context of $f({\cal
G})$ modified theories of gravity. In particular, perturbations in
the energy density and the metric scale factor were introduced, a
linear equation of state, $p(t)=w\rho(t)$, was considered, and
finally the linearized perturbed field equations and the dynamics
of the solutions were analyzed. It was shown that stable modes for
all equation of state parameters $w$ exist, if the parameter
$\alpha$ is chosen appropriately. Thus, as in Refs.~%
\cite{Boehmer:2007tr,Goswami:2008fs,Seahra:2009ft} our results
show that perturbation theory of modified theories of gravity
present a richer stability/instability structure that in general
relativity. Finally, it is of interest to extend our results to
inhomogeneous perturbations in the spirit of Ref.~%
\cite{Seahra:2009ft}, and to include the canonical scalar field
case. Work along these lines is presently underway.

\vspace{-0.5cm}

\acknowledgments We thank Peter Dunsby, Naureen Goheer and Sanjeev
Seahra for useful discussions.

\end{document}